\documentclass[aps,prb,superscriptaddress,reprint]{revtex4-1}

\usepackage{amsmath,amssymb} 
\usepackage{amsmath} 
\usepackage{amsfonts} 
\usepackage{bm} 
\usepackage[pdftex]{graphicx} 
\usepackage{braket} 

\begin{document}

\title{Scan calculation of the density of states: 
real space cluster perturbation theory applied to inhomogeneous Hubbard model in one dimension}

\author{Kaito Matsuki}
\affiliation{Departiment of Physics, The University of Osaka, Machikaneyama, Toyonaka, Osaka 560-0043, Japan}
\author{Chisa Hotta}
\affiliation{Department of Basic Science, The University of Tokyo, Meguro-ku, Tokyo 153-8902, Japan}
\author{Kenichi Asano}
\affiliation{Center for Education in Liberal Arts and Sciences, The University of Osaka, Machikaneyama, Toyonaka, Osaka 560-0043, Japan}
\date{\today}

\begin{abstract}
We present the spectral analysis of a one-dimensional Hubbard model with a parabolic potential,
using a real-space cluster perturbation theory (rCPT) designed to study spatially inhomogeneous
electron systems with strong correlation.
It is a natural extension of the conventional CPT to inhomogeneous cases by computing local Green's functions while averaging over multiple cluster boundaries. 
We find that the local density of states of at each site mirrors that of a homogeneous system with the same local filling.
This insight provides a perspective on spectral evolution in inhomogeneous systems
used to scan the occupancy-dependent features of the homogeneous system, 
making this setup a practical one-shot spectroscopy tool for ultracold atoms in harmonic traps.
\end{abstract}

\maketitle

\section{Introduction}
\label{sec:introduction}
The Hubbard model is a paradigmatic platform for understanding strong correlation effects in electron systems,
capturing essential features of Mott insulators, metallic states, and quantum magnetism.
Since its early exploration, key trends in the study of this model have been made
particularly through cluster perturbation theory (CPT), variational cluster approximation (VCA)\cite{Potthoff2003,Potthoff2004,Aichhorn2006},
and dynamical mean-field theory (DMFT)\cite{Metzner1989,Georges1992,Rozenberg1994,Held2001}.
CPT was first introduced by S\'en\'echal,{\it et. al.} \cite{Senechal2002} 
to study the spectral weight of Hubbard models, 
combining exact diagonalization on small clusters with strong-coupling perturbation theory,
showing applicability to various scenarios, including doping effects and the disappearance
of the Fermi surface. 
VCA extends CPT by including some variational parameters in cluster Hamiltonian\cite{Potthoff2003,Potthoff2004,Aichhorn2006},
and had taken an instrumental role in revealing emergent phases, such as magnetism\cite{Yamada2011}, superconductivity\cite{Sahebsara2006}, and charge or bond orders\cite{Higa2016}. 
Meanwhile, DMFT, which approximates finite-dimensional systems using an infinite-dimensional limit,
was successfully applied to Hubbard models as it captures the local quantum fluctuations exactly
and incorporates frequency-dependent self-energies,
making it highly effective for describing metal-to-Mott-insulator transitions and quasiparticle dynamics
\cite{Georges1992,Rozenberg1994}.
When combined with first-principles calculations, DMFT provides a realistic treatment of electron
interactions in materials such as high-temperature superconductors,
heavy fermion systems, and transition metal oxides \cite{Anisimov1997,Held2001}.
A limitation of DMFT is its neglect of non-local correlations,
which play a crucial role in systems with geometrical frustration.
In that respect, the CPT and VCA provide better insight into correlation effects extending
across multiple lattice sites\cite{Ohashi2006}.
Cluster DMFT (cDMFT) partially addresses this issue and has become a key numerical tool for strongly correlated systems
\cite{Parcollet2004,Maier2005}.
Consequently, a variety of cluster-based methods are now widely employed,
allowing us to choose among CPT, VCA, and cDMFT by balancing computational feasibility with accuracy
when modeling strongly correlated electron systems.
\par
More recently, numerical solvers based on density matrix renormalization group (DMRG) \cite{White1992} and tensor network methods have been developed to compute time evolution \cite{Feiguin2004} and spectral functions \cite{Jeckelmann2002,Gohlke2018,TaoXiang2025}.
These approaches have made low-energy excitations, including full correlation effects,
more accessible, particularly in one (1D) and two-dimensions(2D).
In particular, spectral functions of the 1D Hubbard model have been studied using dynamical DMRG (DDMRG) 
\cite{Jeckelmann2002}
combined with Bethe ansatz \cite{Jeckelmann2004,Kohno2010},
as well as the quantum Monte Carlo method\cite{Zache1998}.
These studies have explicitly identified basic features, such as 
spinon dispersions originating from spin-charge separation effects, 
holon bands, and shadow bands discussed earlier\cite{Senechal2000,Ogata1990,Shiba1991,Penc1996}.
\par
Although the fundamental nature of the excitations in 1D Hubbard model is well established\cite{Penc1996,Bannister2000,Kohno2010},
the model remains a key testing ground for numerical techniques due
to its exact solutions and strong correlation effects.
While extensive studies have explored its spectral properties in homogeneous settings,
real crystalline solids often exhibit spatial inhomogeneity due to external potentials or disorder due to impurities.
Additionally, advances in experimental techniques since the early 2000s have enabled the trapping of Bose gases
in external harmonic potentials \cite{Sedrakian2001},
providing a powerful platform for studying correlation effects in dynamics \cite{Kaneko2022,Kunimi2021}.
Therefore, the ability to systematically analyze the dynamic properties of such inhomogeneous systems is
crucial to characterizing the correlation effects.
\par
In this work, we introduce real-space CPT (rCPT) specifically designed to handle spatially inhomogeneous systems.
Instead of targetting the Green's function in the momentum space, 
rCPT computes local Green's functions. 
By systematically averaging over multiple shifted cluster boundaries, 
this approach effectively removes the artifact from the boundaries between clusters and provides an accurate
description of site-resolved spectral properties. 
Applying rCPT to the 1D Hubbard model, we find that the local density of states (LDOS) at each site,
which depends on site-specific electron occupancy,
closely resembles that of a homogeneous system with the same filling.
This correspondence allows us to interpret spatially varying spectral properties
in terms of occupancy-dependent features of the homogeneous system,
offering new insights into spectral evolution under inhomogeneous conditions. 
Accordingly, the method is used as a scan calculation to obtain the systematic filling-dependence 
of the bulk density of states of the one-particle excitation in correlated systems. 
\par
The paper is structured as follows:
\S.\ref{sec:rcpt} presents the model Hamiltonian and the rCPT framework,
highlighting the development from the conventional CPT.
In \S.\ref{sec:result}, we first apply the CPT to the homogenous 1D Hubbard model to obtain the reference data, 
and then study the rCPT for the 1D Hubbard model with parabolic one-body potential,
where we find that the site-dependent LDOS of the one-particle spectrum
accommodates the DOS of the original translationally symmetric Hubbard model
for a full variation of filling factors. 
The work is summarized in \S.\ref{sec:summary}.

\section{Real space cluster perturbation theory}
\label{sec:rcpt}
\subsection{Preliminaries}
We consider  the one-dimensional (1D) Hubbard of  $N_s$ sites
with hopping $-t<0$ and Coulomb interaction $U\ge 0$, and additionally include a site-dependent one-body potential $v_{i}$.
The Hamiltonian reads
\begin{align}
\mathcal {H}=&-t\sum_{i=1}^{N_s-1} \sum_{\sigma=\uparrow,\downarrow}
    \left( c^{\dagger}_{i+1\sigma}c^{}_{i\sigma}+\text{h.c.} \right) \notag\\
    &+\sum_{i=1}^{N_s} \sum_{\sigma=\uparrow,\downarrow}v_i c^\dagger_{i\sigma}c^{}_{i\sigma}+U\sum_{i=1}^{N_s} c^\dagger_{i\uparrow} c^{}_{i\uparrow}c^\dagger_{i\downarrow} c^{}_{i\downarrow},
\label{eq:ham}
\end{align}
where $c^{}_{i\sigma}$ and $c^\dag_{i\sigma}$ are the annihilation and creation operators of an electron on site $i$, respectively.
When $v_{i}=0$ the system reduces to the homogeneous Hubbard model, where the conventional CPT calculation is performed as a reference to our rCPT. 
In the following calculation, we adopt a parabolic potential,
\begin{equation}
v_i=w\left(\dfrac 2{N_\mathrm{s}}\right)^2\left(i-\dfrac{N_\mathrm{s}}2\right)^2,
\label{eq:vi}
\end{equation}
where $w$ is the potential depth.
\par
Let us express the one-particle retarded Green's function, $\hat G(z)$, as a $N_s\times N_s$ matrix with elements,
\begin{equation}
G_{i,j}(z)=-i\int_0^{+\infty}\left\langle \left\{c^{}_{i\sigma}(s),c^\dag_{j\sigma}\right\}\right\rangle e^{izs}ds,
\label{eq:gr}
\end{equation}
where $z\in {\mathbb C}$ satisfies ${\rm Im} z>0$, $\mathcal O(s)= e^{i(\mathcal H-\mu\mathcal N_e)s} \mathcal O e^{-i(\mathcal H-\mu\mathcal N)s}$ denotes the Heisenberg representation of the operator, and
\begin{equation}
\langle\mathcal O\rangle=\dfrac{\mathrm{Tr}\left(\mathcal Oe^{-\beta(\mathcal H-\mu\mathcal N)}\right)}{\mathrm{Tr}\left(e^{-\beta(\mathcal H-\mu\mathcal N)}\right)},
\label{eq:heisenbrg}
\end{equation}
denotes the expectation value of the operator in the equilibrium at a given inverse temperature, $\beta$, and chemical potential, $\mu$, with $\mathcal N=\sum_{i=1}^{N_s}\sum_\sigma c^\dag_{i\sigma}c^{}_{i\sigma}$ being the electron number operator.
For simplicity, we drop off the spin index, $\sigma$, ensuring that we treat the paramagnetic ground state 
at $\beta\rightarrow\infty$.
\par
In the noninteracting case $(U=0)$, the Green function reduces to
\begin{equation}
\hat G^{(0)}(z)=\left((z+\mu)\hat 1-\hat t\,\right)^{-1},
\end{equation}
where $\hat 1$ is the unit matrix, and $\hat t$ is the one-body Hamiltonian matrix with elements,
\begin{equation}
t_{i,j}=v_i\delta_{i,j}-t\delta_{i-j,1}-t\delta_{i-j,-1}.
\end{equation}
To describe the interaction effects on $\hat G(z)$, we introduce the self-energy matrix, $\hat\Sigma(z)$, so as to satisfy
\begin{equation}
    \hat G(z)=\left((z+\mu)\hat 1-\hat t-\hat\Sigma(z)\right)^{-1}.
   \label{eq:grtot}
\end{equation}
\subsection{CPT Green's function}
Let us divide the system $N_c$ clusters each of size $L$, namely $N_c L= N_s$.
In the following, we label the intra-cluster sites as $i'=1,2,\cdots,L$
and count the whole lattice sites from the left-hand side as $i=(m-1)L+i'$ with the cluster index, $m=1,\cdots, N_c$.
The Hamiltonian of the $m$-th cluster reads
\begin{align}
    \mathcal {H}^{(m)}=&-t\sum_{i'=1}^{L-1} \sum_{\sigma=\uparrow,\downarrow}
        \left( c^{\dagger}_{i'+1\sigma}c^{}_{i'\sigma}+\text{h.c.} \right) \notag\\
        &+\sum_{i'=1}^{L} \sum_{\sigma=\uparrow,\downarrow}v^{(m)}_{i'} c^\dagger_{i'\sigma}c^{}_{i'\sigma}+U\sum_{i'=1}^{L} c^\dagger_{i'\uparrow} c^{}_{i'\uparrow}c^\dagger_{i'\downarrow} c^{}_{i'\downarrow}
        \end{align}
where we relabel the potential as
\begin{equation}
v^{(m)}_{i'}=v_{(m-1)L+i'}.
\end{equation}
The $L\times L$ cluster Green's function matrix, $\hat G^{(m)}$, has the elements,
\begin{equation}
G_{i',j'}^{(m)}(z)=-i\int_0^{+\infty}\left\langle\left\{c^{}_{i'\sigma}(s),c^\dag_{j'\sigma}\right\}\right\rangle e^{izs}ds,
\label{eq:g(m)ij}
\end{equation}
where we redefine $\mathcal O(s)$ and $\langle\mathcal O\rangle$ by replacing $\mathcal{H}$ and $\mathcal{N}$ to $\mathcal{H}^{(m)}$ and $\mathcal N^{(m)}=\sum_{i'=1}^L\sum_\sigma c^\dagger_{i'\sigma}c^{}_{i'\sigma}$.
Resultantly, the cluster self-energy $\hat\Sigma^{(m)}(z)$ is redefined so as to satisfy
\begin{equation}
\hat G^{(m)}(z)=\left((z+\mu)\hat I -\hat t^{(m)}-\hat\Sigma^{(m)}(z)\right)^{-1}.
\end{equation}
where $\hat t^{(m)}$ is the one-body Hamiltoian with elements,
\begin{equation}
t^{(m)}_{i',j'}=v^{(m)}_{i'}\delta_{i',j'}-t\delta_{i'-j',1}-t\delta_{i'-j',-1}.
\end{equation}
\par
The essence of CPT is to approximate the total self-energy as the direct sum of cluster self-energy as
\begin{equation}
\hat\Sigma(z)\approx\bigoplus_{m=1}^{N_c}\hat\Sigma^{(m)}(z).
\end{equation}
Let us introduce the inter-cluster matrix $\hat t^{(c\text{-}c)}=\hat t-\bigoplus_{m=1}^{N_c}\hat t^{(m)}$ whose elements is given as
\begin{equation}
t^{(\mathrm{c\text{-}c})}_{i,j}=-t\left(\delta_{m-n,1}\delta_{i',1}\delta_{j',L}+\delta_{m-n,-1}\delta_{i',L}\delta_{j',1}\right)
\end{equation}
with $i=(m-1)L+i'$, $j=(n-1)L+j'$ and $1\le i',j'\le L$.
Then one can approximate $\hat G(z)$ using $\hat G^{(m)}(z)$ as
\begin{align}
    \hat G(z)&\approx\left(\!(z+\mu)\hat 1-\bigoplus_{m=1}^{N_c}
    \hat t^{(m)}-\hat t^{(c\text{-}c)}-\bigoplus_{m=1}^{N_c}\hat\Sigma^{(m)}(z)
    \!\right)^{-1}\notag\\
    &=\left(\bigoplus_{m=1}^{N_c} \left(\hat G^{(m)}(z)\right)^{-1}-\hat t^{(c\text{-}c)}\right)^{\!-1}
    \notag\\
    &=\left(\bigoplus_{m=1}^{N_c} \hat G^{(m)}(z)\right)
    \left(\hat 1-\hat t^{(\mathrm{c\text{-}c})}\bigoplus_{m=1}^{N_c}\hat G^{(m)}(z)\right)^{-1}.
    \label{eq:CPT Green}
\end{align}
In the actual calculation, we first obtain the ground state $|\Psi^{(m)}\rangle$ 
of a cluster Hamiltonian ${\cal H}^{(m)}$, satisfying 
$({\cal H}^{(m)}-\mu N^{(m)}) |\Psi^{(m)}\rangle = E^{(m)}|\Psi^{(m)}\rangle$, using Lanczos diagonalization. 
The cluster Green's function using $|\Psi^{(m)}\rangle$ and $E^{(m)}$ is given as 
\begin{align}
\hat G_{ij}^{(m)}(z)&= 
\langle \Psi^{(m)} | c_{i\sigma} (z+E^{(m)}+\mu-{\cal H}^{(m)})^{-1} c_{j\sigma}^\dagger |\Psi^{(m)}\rangle \notag\\
&+ \langle \Psi^{(m)} | c_{j\sigma}^\dagger (z-E^{(m)}+\mu-{\cal H}^{(m)})^{-1} c_{i\sigma} |\Psi^{(m)}\rangle, 
\end{align}
for which the continued-fraction form using the Lanczos algorithm is applied\cite{Senechal2008}. 
\par
So far, the formulation is common to the conventional CPT.
The treatment particular to rCPT, which is required to treat the inhomogeneous system, will be introduced shortly in \S.~\ref{sec:recover}.
\subsection{Physical quantities}
The key quantity to treat the inhomogeneous system is the local density of states (LDOS) given as
\begin{equation}
D_{i}(\omega)=-\frac{1}{\pi}\mathrm{Im}G_{i,i}(\omega+i\delta)
\label{eq:ldos}
\end{equation}
where $\omega$ is the one-particle energy and $\delta$ is the positive constant, which is taken as $\delta/t = 0.1$ in our study.
This quantity represents the energy distribution that the electron at site $i$ carries and its 
average over $i$ gives the standard density of states, $D(\omega)=N_s^{-1}\sum_i D_i(\omega)$. 
\par
Another key quantity is the electron number at site $i$,
\begin{align}
n_i&=\Big\langle \sum_\sigma c_{i\sigma}^\dag c^{}_{i\sigma}\Big\rangle
=2\int_{-\infty}^{0}D_i(\omega)d\omega\notag\\
 &=-\dfrac 2\pi\int_{-\infty}^0{\rm Im}G_{i,i}(\omega+i\delta)d\omega.
\end{align}
Although this expression is rigorous, the large variance of Green's function near the real axis makes it practically difficult
to evaluate it accurately.
We thus modify the path of integration to the one along the imaginary axis,
and to guarantee this treatment,
we make use of the convergence of $G_{i,i}(z)-G_{i,i}^{(0)}(z) \sim 1/z^2$ at
$|z|\rightarrow+\infty$, much faster than $G_{i,i}(z)\sim 1/z$ itself.
Then we obtain
\begin{align}
n_i&=n_i^{(0)}
-\dfrac 2\pi\mathrm{Im}\int_{-\infty}^0\left(G^{}_{i,i}(\omega+i\delta)-G_{i,i}^{(0)}(\omega+i\delta)\right)d\omega
\notag\\
    &=n_i^{(0)}+\dfrac 2\pi\int_0^{+\infty}\mathrm{Re}\left(G^{}_{i,i}(ix)-G_{i,i}^{(0)}(ix)\right)dx
    \label{eq:numerical_ni}
\end{align}
with
\begin{align}
n_i^{(0)}&=-\dfrac 2\pi\int_{-\infty}^0\mathrm{Im}G_{i,i}^{(0)}(\omega+i\delta)d\omega\notag\\
&=-\dfrac 2\pi\int_{-\infty}^0\mathrm{Im}\left(\sum_{\alpha}\dfrac{|u_{\alpha,i}|^2}{\omega-\epsilon_\alpha+\mu+i\delta}\right)d\omega\notag\\
&=2\sum_{\epsilon_\alpha<\mu}|u_{\alpha,i}|^2,
\end{align}
where $u_{\alpha,i}$ denotes the $i$th element of the eigenvector of $\hat t$ with the eigenenergy $\epsilon_\alpha$.
\par
At $U\ne 0$, the Hartree decomposition of the on-site Coulomb interaction term yields,
\begin{align}
    U\sum_i c^\dag_{i\uparrow}c^{}_{i\uparrow}c^\dag_{i\downarrow}c^{}_{i\downarrow}
&\approx
U\sum_i\Big(\langle c^\dag_{i\uparrow}c^{}_{i\uparrow}
    \rangle c^\dag_{i\downarrow}c^{}_{i\downarrow}
   + \langle c^\dag_{i\downarrow}c^{}_{i\downarrow}\rangle
    c^\dag_{i\uparrow}c^{}_{i\uparrow}
\notag \\
& \rule{10mm}{0mm}-\langle c^\dag_{i\uparrow}c^{}_{i\uparrow}\rangle
\langle c^\dag_{i\downarrow}c^{}_{i\downarrow}\rangle\Big)
\end{align}
By including this contribution, one can define the effective one-body potential as
\begin{equation}
    v_i^\mathrm{(eff)}=v_i+\dfrac U2 n_i\label{eq:pot_eff},
\end{equation}
which includes the correlation effect beyond the naive Hartree approximation, since we evaluate $n_i$ by Eq.~(\ref{eq:numerical_ni}).
\begin{figure*}
    \centering
    \includegraphics[width=17cm]{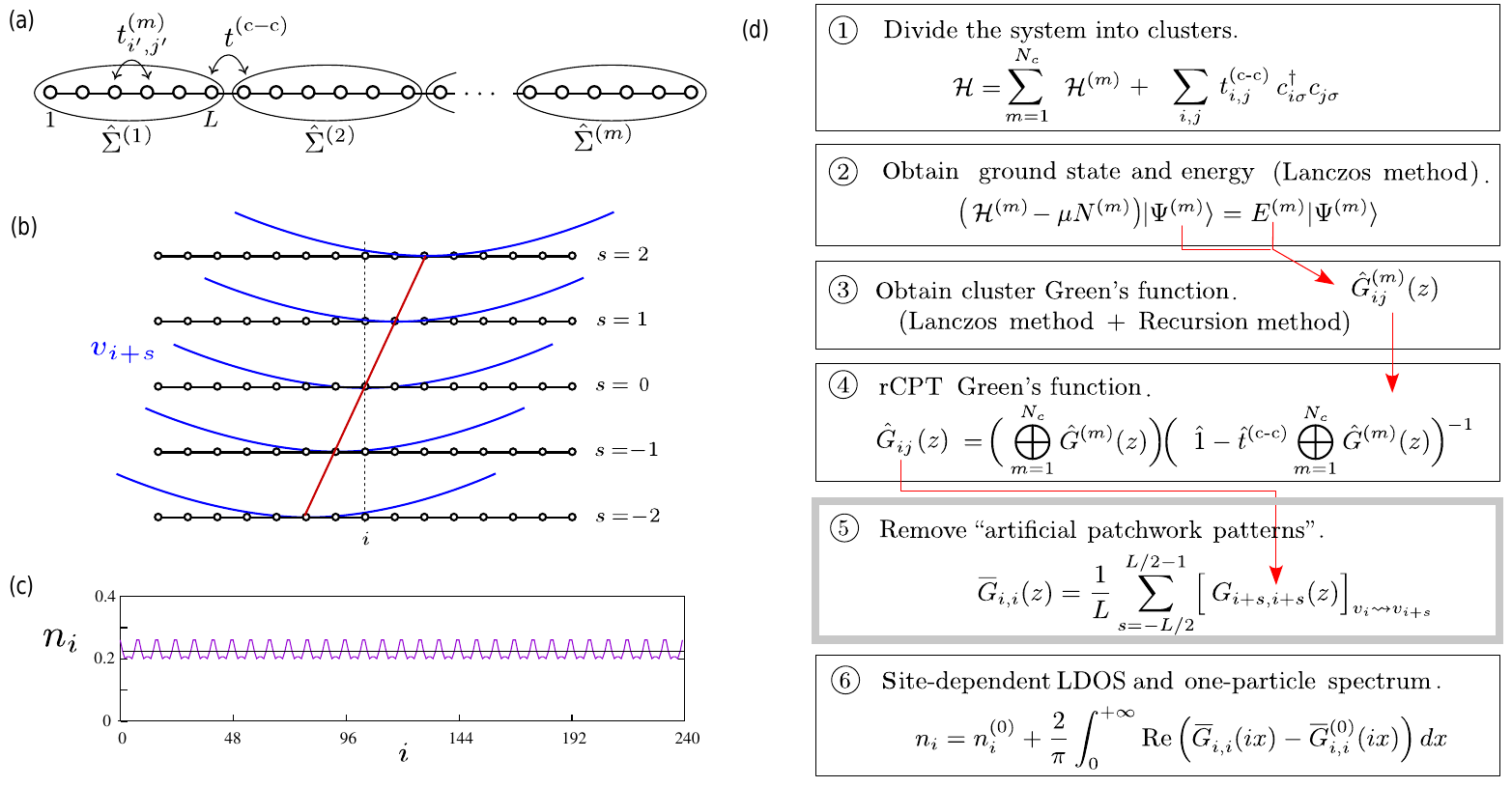}
    \caption{Schematic illustration of (a) CPT/rCPT and
(b) the treatment of recovering ``translational invariance" given in Eq.(\ref{eq:recover}).
(c) the electron number $n_i$ at site $i$ obtained before and after applying Eq.(\ref{eq:recover}),
where we take $w=15$,$U=2$, $N_s=240$ with PBC and $L=8$.
(d) Flow diagram of rCPT.
}
    \label{f1}
\end{figure*}
\subsection{rCPT procedure}
\label{sec:recover}
In CPT, the hoppings between clusters and those within a cluster are treated inequivalently by construction (see Fig.~\ref{f1}(a)).
The consequence is clearly observed when CPT is applied to the spatially homogeneous case, $v_i=0$; the original single-site periodicity of the system is lost, and the electron number and LDOS exhibit site-dependent oscillations with a period of $L$, which is not of a physical origin.
The spatially inhomogeneous system also suffers similar "artificial patchwork patterns"; the physical quantities deviate from those expected for a smooth profile of $v_i$.
\par
Here, we propose real-space CPT (rCPT) which adopts the following treatment to the local Green's functions, $G_{i,i}(z)$, to remove such ``artificial patchwork patterns''.
While shifting the joint position of the clusters by $s=-L/2,\cdots, 0,1,\cdots , L/2-1$ relative to one another, 
we calculate $G_{i,i}(z)$ and redefine it as their average.
For the system with open boundary condition (OBC), shifting the joint position changes the size of the cluster at the system edges.
Therefore, we instead shift the potential as $v_i \leadsto v_{i+s}$ and consider $G_{i+s,i+s}(z)$ in the inner part of the system as shown schematically in Fig.~\ref{f1}(b). 
Then we evaluate the averaged local Green's functions as
\begin{equation}
  \bar G_{i,i}(z)=\dfrac{1}{L}\sum_{s=-L/2}^{L/2-1} \Big[ G_{i+s,i+s}(z)\Big]_{v_i\leadsto v_{i+s}} 
\label{eq:recover}
\end{equation}
This averaging procedure does not work near the system edge where $i+s$ sticks out of the system.
In our case, however, it does not matter because we choose the potential depth, $w$, enough large to deplete electrons 
completely from the edge.
\par
The conventional CPT, which is specialized to a homogeneous system of $v_i=0$ with periodic boundary condition (PBC), 
removes the ``artificial patchwork patterns'' by omitting the off-diagonal element of the Green's function in the momentum representation:f
\begin{align}
    G(k, k'; z)=\frac{1}{N_\mathrm{s}}\sum_{i,j}G_{i,j}(z)e^{-iki}e^{ik'j}\approx \delta_{k,k'}G(k,z),
\end{align}
where we define $G(k,z)=G(k,k;z)$, and $k,k'\in (-\pi,\pi]$ are integral multiples of $2\pi/N_s$.
Then, the inverse Fourier transformation yields
\begin{align}
    G_{i,i}(z)&=\dfrac 1{N_s}\sum_k G(k,z)=\dfrac 1{N_s}{\rm Tr}\hat G(z),
    \label{eq:conventional}
\end{align}
recovering the site-independence.
One can here adopt Eqs.~(\ref{eq:Gkk0}) and (\ref{eq:Gkk}) in Appendix~\ref{app:cptvsrcpt} to evaluate $G(k,z)$, instead of direct Fourier transformation of $G_{i,j}(z)$.
When applying rCPT to the homogenous cases, Eq.(\ref{eq:recover}) gives the same result as
\begin{equation}
 \bar G_{i,i}(z)=\dfrac 1L\sum_{i'=1}^{L}G_{(m-1)L+i',(m-1)L+i'}(z)=\dfrac 1{N_s}{\rm Tr}\hat G(z),
 \label{eq:rCPT_homo}
\end{equation}
since $G_{(m-1)L+i',(m-1)L+i'}(z)$ is independent of $m$.
One can thus consider rCPT as a natural extension of the conventional CPT to the inhomogeneous cases.
\par
Figure~\ref{f1}(c) shows the electron number $n_i$ at site $i$ obtained by rCPT at $v_i=0$ and $N_s=240$ with PBC and $L=8$, before and after applying Eq.(\ref{eq:recover}) which removes the artificial oscillation of $L$-site periodicity. 
The similar comparison of the data before and after applying Eq.(\ref{eq:recover}) 
for the inhomogeneous system will be presented shortly in Fig.~\ref{f5}. 
In Fig.~\ref{f1}(d) we show the flow diagram outlining the rCPT processes. 

\section{results }
\label{sec:result}
We first show the results of the $v_i=0$ case in \S.\ref{sec:res_homogeneous} as a reference.
Then in \S.\ref{sec:res_parabollic}
we examine how the electronic state and the excitation spectrum change when the parabolic potential
is considered, and discuss the physical implication of the results.

\begin{figure*}
    \centering
   \includegraphics[width=16cm]{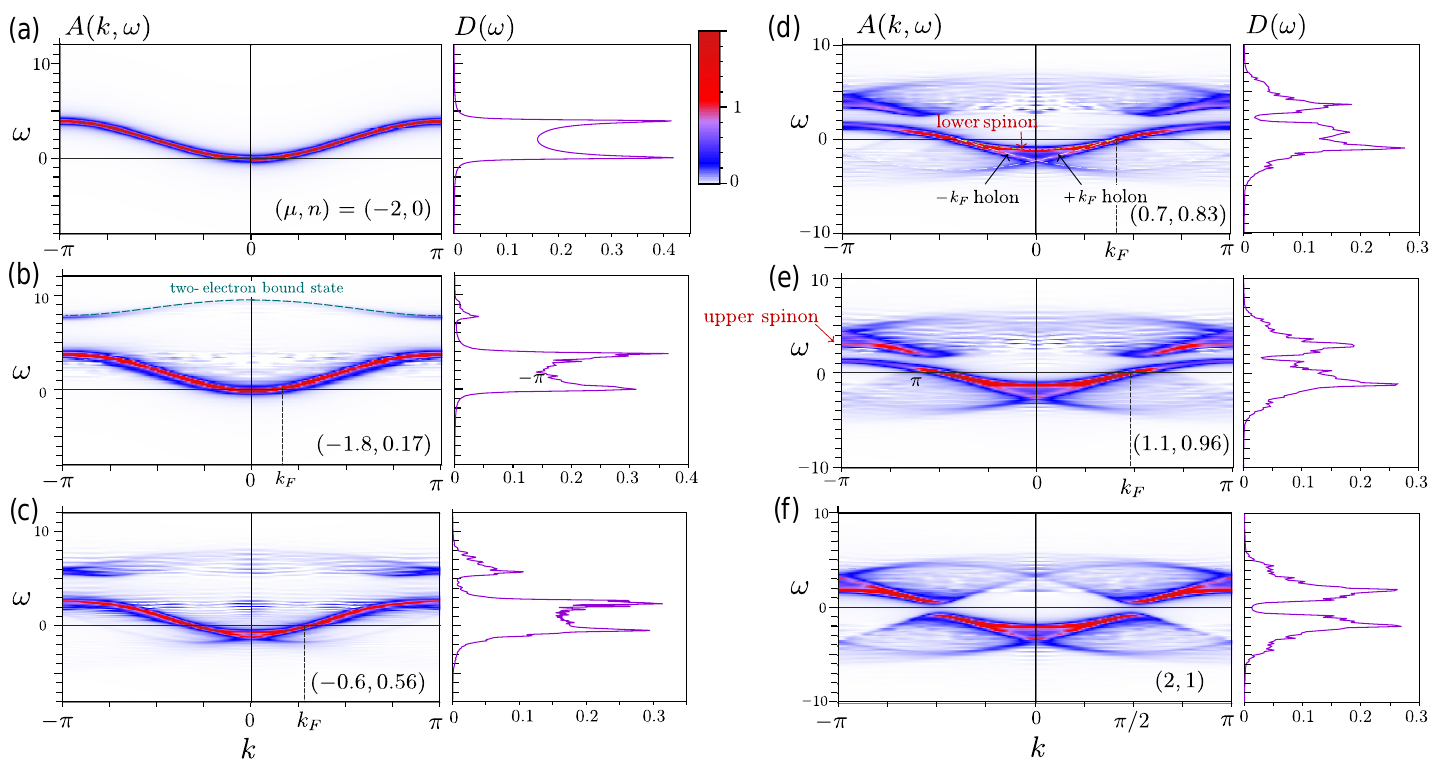}
    \caption{One particle spectrum, $A(k,\omega)$, obtained by the conventional CPT using $L=12$ with PBC,
and the corresponding density of states (DOS), $D(\omega)$.
We take for (a)-(f) as $(\mu,n)=(-2,0), (-1.8, 0.17), (-0.6,0.56), (0.7,0.83),
(1.1,0.96), (2,1)$, respectively.
The energy dispersions are labeled as $\pm k_F$-holon, and upper/lower spinon bands. 
In panel (b) we plot Eq.(\ref{eq:doublon}) that exactly represent the two-electron bound state dispersion
in blue broken line, whose details are given in Appendix~\ref{app:lowd}.
}
    \label{f2}
\end{figure*}
\subsection{Spatially homogeneous case $v_i=0$}
\label{sec:res_homogeneous}
We first apply the conventional CPT to the homogeneous case ($v_i=0$) as a reference system to be compared with the inhomogeneous case.
Using the cluster of $L=12$, we evaluate the one-particle spectra,
\begin{equation}
A(k,\omega)=-\dfrac 1\pi\mathrm{Im}G(k,\omega+i\delta).
\end{equation}
The DOS (LDOS) obtained as its average over $k\in (-\pi,\pi]$.
When we apply PBC of $N_s$ sites letting $k$ be integral multiples of $2\pi/N_s$, the DOS strictly agrees with the one obtained by rCPT, as shown in Eq.~(\ref{eq:rCPT_homo}).
\par
Figures~\ref{f2}(a)-\ref{f2}(f) show the results of $U=4t$ for several filling factors, $n\le 1$.
At zero filling, when the chemical potential is set to $\mu=-2t$, we have $n=0$ where $A(k,\omega)$ reproduces the noninteracting dispersion, $\omega=-2t\cos k-\mu$, starting from the chemical potential level ($\omega=0$), and the corresponding DOS follows
\begin{equation}
D(\omega)=\dfrac 1{\pi\sqrt{4t^2-(\omega+\mu)^2}},
\end{equation}
showing van Hove singularities at the band edges, $\omega=0$ and $4t$. 
Similar cosine dispersion is maintained at $n= 0.17$ ($\mu=-1.80t$), which we call ``main band", 
and additionally, there appears a satellite band of a width $\sim 4t^2/U$ 
with its lower edge starting from $\omega\sim 8t$ and $k\sim \pi$. 
This gives a cosine-like dispersion coming from the two-electron bound state, 
showing good agreement with the analytically exact solution (broken line in panel (b)) 
for the dilute limit (two electrons) given in Appendix \ref{app:lowd}, Eq.(\ref{eq:2electron}). 
\par
In Figs.~\ref{f2}(c) and \ref{f2}(d), the main band at $|k|\le k_F$ and $\omega<0$ 
branches into two parts; the upper branch is called the spinon band (short-broken line)
that follows $\sim -\pi J/2 \cos k$ with $J=4t^2/U$ at large enough $U$. 
The lower branch at $0\le k \le k_F$ is a holon band,
that continues toward $-3k_F \le k \le 0$ which is 
called shadow band (see the long-broken line, $-k_F$-holon). 
The holon-shadow bands form a cosine structure with its wave number shifted by $-k_F$, 
and there is another one with a $+k_F$ shift. 
These structures are known from the spin-charge separation and the Luttinger liquid theory.
\par
The weight of the main band is gradually transferred from $|k|>k_F$ and $\omega>0$ to $|k|<k_F$ and $\omega<0$. 
In Fig.~\ref{f2}(e) the satellite band on the upper part at $k\sim\pi$ 
starts to branch, whose lowest part becomes the upper spinon band. 
Near half-filling (see Figs.~\ref{f2}(e) and \ref{f2}(f)), the main band at $\omega>0$
loses its weight which transforms to the upper spinon band. 
The bands at $\omega>0$ become symmetric at half-filling about $\pi$ rotation 
with respect to $(k, \omega) = (\pi/2, 0)$. 
\par
While these results basically reproduce those given in Refs.[\onlinecite{Jeckelmann2004}] and [\onlinecite{Kohno2010}] based on
DDMRG and Bethe ansatz, the CPT here
both quantitatively and qualitatively display the way how the DOS or the spectrum evolves
in more detail based on the far simpler and far less costly framework.
The rCPT we see shortly give comparably accurate results.

\begin{figure*}
    \centering
   \includegraphics[width=16cm]{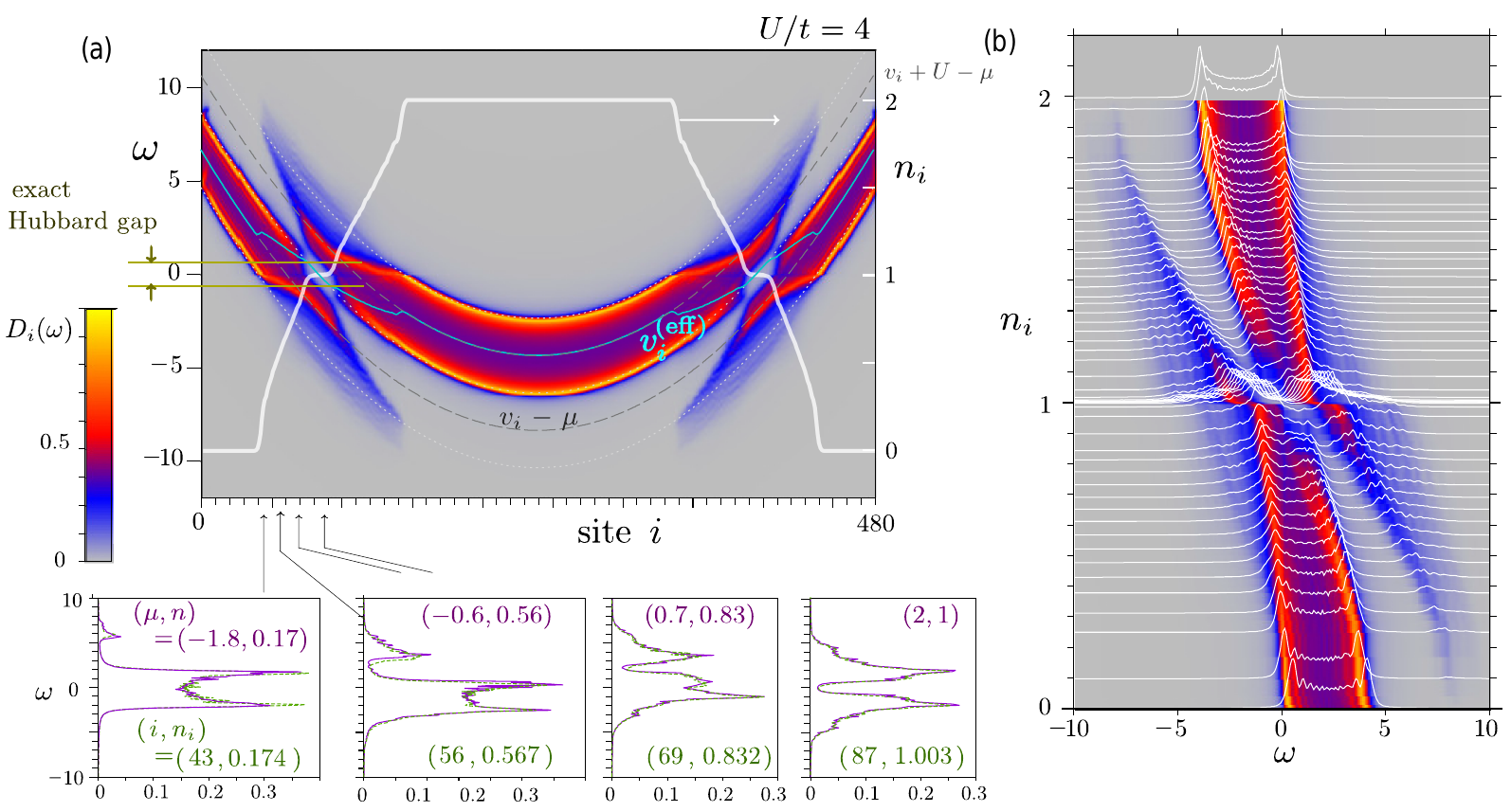}
    \caption{Local density of states, $D_i(\omega)$ obtained by rCPT with $v_i$
    in Eq.(\ref{eq:vi}) with $w/t=15$ and at $U/t=4$ for $N_s=480$ and $L=12$.
    Panel (a) shows the density plot of $D_i$ on the $i$-$\omega$ plane,
    together with $n_i$ (right vertical axis).
    We plot $v_i^{\text{(eff)}}$ (solid line) and $v_i+U-\mu$, $v_i-\mu$(broken lines),
    $v_i+U-\mu\pm 2t$, $v_i-\mu\pm 2t$ (dotted line) for the guide to the eye.
    We plot the location of the Hubbard gap of the Bethe ansatz solution, $\Delta=1.2867$ to compare with our data. 
    The lower insets are those at several fixed sites, $i=43, 56, 69, 87$,
    where we plot together in broken lines the DOS, $D(\omega)$, of the homogeneous systems
    with $n$ being similar to $n_i$.
    Panel (b) shows $D_i$ both in the density plot and a series of lines, each origin being set to $n_i$ on that site.
}
    \label{f3}
\end{figure*}
\begin{figure*}[t]
    \centering
   \includegraphics[width=18cm]{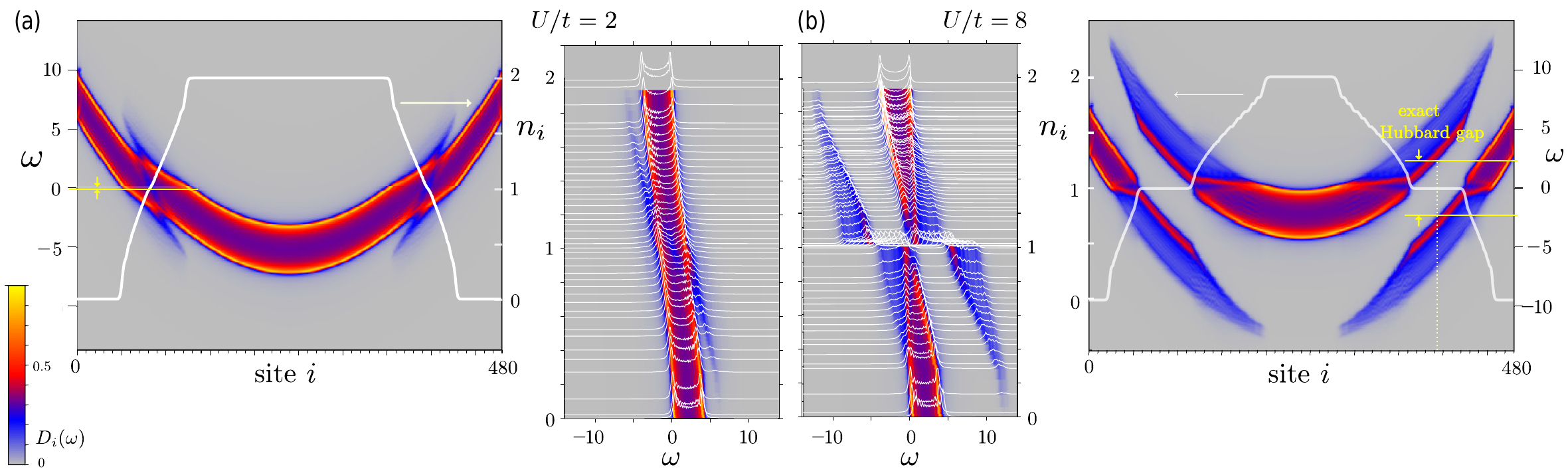}
    \caption{Local density of states, $D_i(\omega)$ obtained by rCPT with $v_i$
    in Eq.(\ref{eq:vi}) with $w=15$ at (a) $U/t=2$ and (b) 8, using $N_s=480$ and $L=12$.
    These data are written in the same scale as Fig.~\ref{f3}.
    We plot the location of the Hubbard gap of the Bethe ansatz solution, $\Delta=0.1727$ ($U/t=2$) and $4.6795$ ($U/t=8$)
    to compare with our data.
}
    \label{f4}
\end{figure*}
\subsection{Parabollic potential}
\label{sec:res_parabollic}
We now consider the Hubbard model with a parabolic potential $v_i$ given in Eq.(\ref{eq:vi}).
The parameters are taken as $N_s=480$, $w=15$ where we set the chemical potential as
\begin{equation}
\mu=\frac{1}{2}(cw + U),
\end{equation}
with constant $c\in {\mathbb R}$.
By taking $c= 0.8-0.9$, the effective chemical potential at site $i$ given as
$\mu_i^{\text{(eff)}}=\mu -v_i$ increases from the
system edges toward the center, crossing $\mu_i\sim 0$ somewhere in between,
and one can find both the fully occupied and fully empty sites within the system.
In the present case, we set $cw =v_{18}$,
for which we can adjust the location of sites that have a $n_i\sim 1$
common to different values of $U$, which enables us to compare the results for different $U$.
For these choices, the average electron density for a given $\mu$ will be in between $1.2-1.3$.
\par
We present the LDOS given by Eq.(\ref{eq:ldos}) and the particle density at site $i$ calculated using Eq.(\ref{eq:numerical_ni}).
We first focus on the results at $U/t=4$ in comparison with the $A(k,\omega)$ and $D(\omega)$ we saw in Fig.~\ref{f2}. 
Figure~\ref{f3}(a) shows the density plot of LDOS on the plane of site index $i$ and $\omega$,
Here, we find that the center of the spectrum at each site, 
namely the center of the strong intensity part about $\omega$, 
is given by the function,
\begin{equation}
v_i^{\text{(eff)}}=v_i+\frac{U}{2}n_i,
\label{eq:vieff}
\end{equation}
shown in a solid line as a guide to the eye.
\par
In the lower panels of Fig.~\ref{f3}(a), we show $D_i(\omega)$ as a function of $\omega$ 
for $i=43, 56, 69, 87$, 
and for each, we plotted together $D(\omega)$ of the homogeneous reference system 
having the similar filling factors as $n_i$ where the origin of $\omega$ for $D(\omega)$ 
is shifted by $v_i^{\text{(eff)}}-\mu$. 
Importantly, these two show excellent agreement. 
This indicates that the system with $v_i\ne 0$ carries the ``full set" of
information of the bulk one-particle density of states 
of the corresponding homogeneous system $v_i=0$ as a function of filling factor $n_i$.
\par
To take a closer look at how the spectrum evolves, we also plot $v_i+U-\mu$ and $v_i-\mu$ 
as broken lines in Fig.~\ref{f3}(a), 
and their range of bandwidth, $v_i+U-\mu \pm 2t$ and $v_i-\mu \pm 2t$, in dotted lines (part of them overlap). 
When the potential excludes the electron occupancy near the edges of the system, 
i.e. $n_i\sim 0$ at $i\lesssim 18$, $v_i^{\text{(eff)}}-\mu$ coincides with $v_i-\mu$,
whereas at the system center where the electrons doubly occupy the site,
$v_i^{\text{(eff)}}-\mu$ shows a Hartree-shift and coincides with $v_i+U-\mu$.
Correspondingly, $n_i$ shows a plateau at $n_i=0$ and $2$ near the edge and the center, respectively,
while there appears another plateau $n_i=1$ at around $i\sim 88$ where 
$v_i^{\text{(eff)}}-\mu \sim 0$.
This plateau indicates the presence of a Mott gap in LDOS.
We show the location of the Hubbard gap at $U/t=4$ of the exact Bethe ansatz solution in the figure\cite{Schulz1993},
which agrees with the range between the peaks of the LDOS at $i\sim 85$. 
Notice that near the plateau at $i\le 240$, $v_i^{\text{(eff)}}-\mu$ is neither a smooth
nor a simple increasing function of $i$, 
which is because $v_i-\mu$ and $Un_i/2$ are the decreasing and increasing function of $i$, respectively.
\par
With the above considerations in mind, we replot in Fig.~\ref{f3}(b) the same data as panel (a)
according to the ``filling factor", $n_i$, instead of the site index. 
The origin of each spectrum line is set to $n_i$, 
to show how the structure of the spectrum evolves in varying electron filling, 
and the intensity plot is shown together for clarity. 
When $n_i=0$, which corresponds to $i\lesssim 20$, the LDOS exhibits the same shape as that for $U=0$,
showing van Hove singularity peaks at the edges of the main band.
As $n_i$ increases slightly, there starts to appear weak structures at higher energies 
$\omega\sim 8t$ in addition to the main band,
which corresponds to the satellite band representing the doublon states.
\par
When $n_i$ approaches 1, among the two van Hove singularity peaks originally associated with the main band, 
the upper peak at around $v_i-\mu + 2t$ of $n_i<1$ weakens and almost fades out, 
and its intensity shifts to the upper subband. 
The lower van Hove peak continues and at the opening of the Mott gap at $i\sim 88$ 
shows a rather abrupt shift to the lower $\omega$ and transitions into the upper part 
of the lower Hubbard band. 
Meanwhile, the upper subband starting at $\omega\sim 8t$ grows and connects to the upper
van Hove singularity at around $v_i+ U + 2t -\mu$ of $n_i>1$. 
\par
This behavior can be explained in reference to Fig.~\ref{f2}(e); 
the main band above the chemical potential level at $|k| \ge k_F$ starts to 
transfer its weight to the upper Hubbard band and fades out in opening a gap, 
which corresponds to the center line among the three structures in $D_i(\omega)$ of Fig.~\ref{f3}(b). 
At $|k| \le k_F$, the main band branches to spinon and holon and increase its intensity, 
and the $k\sim 0$ part becomes the main intensity of the lower Hubbard band, 
which is observed in Fig.~\ref{f3}(b) as the merging of 
the two van Hove red peaks merging at $n_i\rightarrow 1$. 
The holon band of $2k_F$ periodicity extends to the lower part of $\omega$ and forms a shadow band 
with a long tail \cite{Penc1996}, which forms the bottom of the lower Hubbard band. 
How the peak structure evolves and relates to each other when varying $n_i$ is
observed using a single shot of calculation under the parabolic potential.
\par
To confirm that the functionality of putting parabolic potential works
for other parameters, we examine the case of $U/t=2$ and $8$ as shown in Fig.~\ref{f4}.
When $U/t=2$, the Mott gap is finite but small ($\Delta= 0.1727$ for the Bethe ansatz solution)
which is almost invisible in the figure, in which case $n_i$ shows a slight
anomaly at $n_i=1$, which is at $i=87$, but is not enough to form a clear plateau.
Otherwise, due to small interactions, the
the main structure of the LDOS consists of sharp peak structures
originating from van Hove singularities at the edges of the main band similar to the
noninteracting ones, with only small satellite peaks appearing.
\par
When $U/t=8$, the system shows a very stable and wide plateau at $n_i=1$.
The Mott gap of the LDOS at the center of the plateau agrees with the Bethe ansatz solution
$\Delta=4.6795$ shown in the figure.
The larger interactions allow various features to be more distinctly visible.
There are two main structures in the spectrum; 
the main band accompanying the shadow band and the satellite band.
Near half-filling, the two van Hove peaks merge within the main band but 
the width including the shadow part does not change much, 
whereas the satellite band increases its width. 
In approaching from both sides, $n_i\rightarrow 1+$ and $n_i\rightarrow 1-$,
the two main structures meet with those of the two on the other side of the filling,
where we find how the Hubbard bands at $n_i=1$ are constructed. 
The particle-hole symmetric pairs of these origins are observed.

\subsection{Mapping between site location and $\mu$}
We demonstrated that the local density of states $D_i(\omega)$ at site $i$, 
calculated using rCPT under a parabolic on-site potential $v_i$, 
closely matches the bulk density of states of a homogeneous system with the same local filling $n_i$. 
In this comparison, $n_i$ is treated as an intrinsic parameter, and $D_i(\omega)$ is plotted as a function of $n_i$.
Alternatively, one may use the effective chemical potential $\mu_{\rm b} \equiv \mu - v_i$, 
which is expected to produce the same filling $n_i$ in the corresponding homogeneous system.
\par
\begin{figure}[t]
    \centering
   \includegraphics[width=8cm]{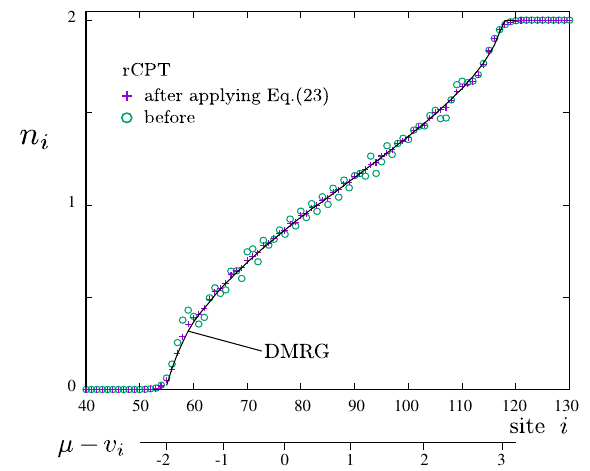}
    \caption{Particle density of a homogeneous Hubbarad model at $U/t=1$ as a function of chemical potential $\mu_b$ 
obtained by the scale-free type of DMRG (solid line), where we take $\mu_b=\mu-v_i$ for $i=55-120$. 
The DMRG data is compared with the rCPT data with a parabollic potential, $w=15$, and $\mu=6.915$ 
as a function of site $i$, where we show two series of data before/after (circle/cross) 
applying Eq.(\ref{eq:recover}) to recover the cluster-based ``translational invariance". 
}
    \label{f5}
\end{figure}

To test this interpretation, we independently calculate the particle density as a function 
of $\mu_{\rm b}$ in the homogeneous Hubbard model using the size-scaling-free DMRG method \cite{Hotta2012,Hotta2013}. 
This method yields bulk particle densities with high accuracy, within $10^{-3}-10^{-4}$ of the exact values. 
The DMRG result for $U/t = 1$ is shown in Fig.~\ref{f5} as a solid line, where the horizontal axis is indexed by site $i$, 
which maps to $\mu_{\rm b} = \mu - v_i$. 
The rCPT data, shown as cross symbols, exhibits good agreement with the DMRG curve. 
These results support the conclusion that, at each site $i$, 
the local particle density $n_i$ corresponds to that of a homogeneous system at chemical potential $\mu_{\rm b} = \mu - v_i$, 
and that the local spectral function $D_i(\omega)$ represents the one-particle excitation spectrum of 
that homogeneous system.
\par
Figure~\ref{f5} also includes the unprocessed rCPT data prior to applying Eq.(\ref{eq:recover}) in circles. 
We find that the process effectively smooths the rCPT data and brings it into quantitative agreement 
with the DMRG result, within a typical deviation of order $10^{-3}$. 

\section{summary and discussion}
\label{sec:summary}
We studied the spectral property of the spatially inhomogeneous 
one-dimensional Hubbard model in the presence of a harmonic trap, 
using a real-space extension of Cluster Perturbation(rCPT) 
which is designed as a natural extention of conventional CPT to address inhomogeneous correlated systems. 
\par
Since CPT is a method that divides the system into clusters,
a naive approach would lead to "patchwork patterns", 
which can have a non-negligible artifact on the calculation results. 
In the conventional CPT, this issue is avoided by 
first obtaining the Green's function in momentum space, 
and then restoring translational symmetry by discarding off-diagonal elements in momentum representation.
However, in an inhomogeneous system, this approach is not applicable
because the wave number is no longer a quantum number. 
In our rCPT, the local Green's functions are computed, while they are averaged over the states
for which the cluster boundaries are shifted. 
\par
We applied the rCPT to a 1D Hubbard model with a parabolic potential. 
To verify the effectiveness of rCPT, we first prepared the reference result by performing 
the conventional CPT for a homogeneous 1D Hubbard model. 
In rCPT, the local density of states and electron occupancy $n_i$ at each site are obtained. 
Both quantities vary depending on their real-space location, 
which turned out to store the intrinsic information about the original homogeneous Hubbard model;
The profile of the local density of states at site $i$ matches with high accuracy the reference data, 
i.e. the density of states of a homogeneous system of the same system size having the filling factor $n=n_i$. 
This finding allows us to interpret the site dependence of the local density of states in terms of
the occupancy dependence of the density of states in a homogeneous system. 
By arranging the local density of states across different sites,
a single rCPT calculation can provide results corresponding to a continuous variation in occupancy,
making it a practical computational method. 
\par
Through comparison with DMRG, we conclude that the site-dependent quantities in the inhomogeneous system 
are equivalent to those of a homogeneous system obtained for the chemical potential 
with a shift by the corresponding potential value. 
Such aspect has future persectives; 
When applied to ultracold atoms in optical traps \cite{Jaksch2005,Bloch2008}, 
it becomes possible to experimentally track the time-dependent particle density and local density of states, 
whose Fourier transform yields $D_i(\omega)$, and the equilibrium value after thermalization corresponds to $n_i$. 
Therefore, by controlling the potential profile, one can effectively perform a one-shot spectroscopy in laboratory 
to obtain a variants of quantities of a corresponding homogeneous system. 
These experimental setup may also be traced numerically in theories. 
While systems like Rydberg atoms involve longer-range inter-site interactions, 
which may introduce some limitations to the accuracy of rCPT, the broader insight remains valid: 
inhomogeneous systems encode bulk spectral information over a variation of system parameters. 
It is not restricted to rCPT and can be extended to other models and computational methods. 
\par
Introducing inhomogeneity into model parameters as a numerical framework has been successfully implemented
in various methods and models.
For example, in DMRG, the transfer integral or chemical potential of the $t$-$t'$-$J$ model
was gradually varied across a cylindrical system, enabling the determination of the phase diagram
along an entire parameter line in a single calculation, known as a scan calculation \cite{Jiang2021}.
Similarly, sine-square deformation (SSD) modulates the Hamiltonian spatially while preserving the essential physics
of the original model \cite{Gendiar2009,Maruyama2011}, allowing precise measurement of magnetization curves
in frustrated magnets \cite{Hotta2012,Hotta2013,Nishimoto2013}.
In the classical Ising model, the same deformation, namely site-dependent modifications of Hamiltonian parameters
have enabled Monte Carlo simulations to simultaneously explore quasi-equilibrium states across
different effective temperatures \cite{Hotta2021},
conceptually similar to scan calculations.
These approaches share common aspects with the present rCPT method,
where the excitation spectrum is evaluated locally in a system with smoothly varying potentials,
effectively functioning as a scanning spectroscopy technique.

\begin{acknowledgments}
This work is supported by KAKENHI Grant No. 21K03440 
from JSPS of Japan. 
\end{acknowledgments}

\appendix

\section{$G(k,z)$ in conventional CPT}
\label{app:cptvsrcpt}
In the homogeneous case, the cluster Green's function matrix, $G^{(m)}(z)$, in
Eq.(\ref{eq:g(m)ij}) becomes independent of $m$, which we denote as $G^{(c)}(z)$.
Let us define a $N_s\times N_s$ unitary matrix $\hat U$ denoting Fourier transformation in units of $L$ as
\begin{equation}
U_{(\tilde k,i'),j}=\dfrac{e^{-i\tilde k n L}}{\sqrt{N_c}}\delta_{i',j'},
\end{equation}
where the wave number, $k\in (-\pi/L,\pi/L]$, is integral multiples of $2\pi/N_s$ taking $N_c$ discrete values, $n=1,2,\cdots,N_c$ is the cluster index, $i',j'=1,2,\cdots,L$ are intracluster site indices, and $j=(n-1)L+j'=1,2,\cdots,N_s$ is the whole site index.
We should define $\tilde k$, not in the original Brillouin zone, $(-\pi,\pi]$, but in the folded one, $(-\pi/L,\pi/L]$, since CPT introduces artificial $L$-site periodicity by dividing the system into clusters.
\par
The $L$-site periodicity results in the following block diagonalization,
\begin{align}
&\hat U\big(\bigoplus_{m=1}^{N_c}\hat G^{(c)}(z)\big)\hat U^\dag=\bigoplus_{\tilde k}\hat G^{(c)}(z), \\
 &\hat U\hat t^{(c\text{-}c)}U^\dag=\bigoplus_{\tilde k}
 \hat t^{(c\text{-}c)}(\tilde k)
\end{align}
with
\begin{equation}
    \hat t_{i'j'}^{\mathrm{(c\text{-}c)}}(\tilde k)=-t\big(e^{-i\tilde kL}\delta_{i_1',1}\delta_{i_2',L}+e^{i\tilde kL}\delta_{i_1',L}\delta_{i_2',1}\big).
\end{equation}
Then Eq.~(\ref{eq:CPT Green}) yields
\begin{align}
&\hat U\hat G(z)\hat U^\dagger =\bigoplus_{\tilde k}\hat G(\tilde k,z)\\
&\hat G(\tilde k,z)=\hat G^\mathrm{(c)}(z)\Big(\hat 1-t^{(\mathrm{c\text{-}c})}(\tilde k)\hat G^\mathrm{(c)}(z)\Big)^{-1},
\label{eq:Gkk0}
\end{align}
and thus
\begin{align}
G(k,z)
=&\frac{1}{N_\mathrm{s}}\sum_{i,j}G_{i,j}(z)e^{-ik(i-j)}\notag\\
=&\frac{1}{N_\mathrm{s}}\sum_{m,n}\sum_{i',j'}G_{(m-1)L+i',(n-1)L+j'}(z)\notag\\
&\hspace{2cm}\times e^{-i\tilde k(m-n)L}e^{-ik(i'-j')}\notag\\
=&\frac{1}{L}\sum_{i',j'}\left(U^\dag\hat G(z)U\right)_{(\tilde k,i'),(\tilde k,j')}e^{-ik(i'-j')}\notag\\
=&\frac{1}{L}\sum_{i',j'}G_{i',j'}(\tilde k,z)e^{-ik(i'-j')},
\label{eq:Gkk}
\end{align}
where we reduce $k\in (-\pi,\pi]$ to $\tilde k\in (-\pi/L,\pi/L]$ so as to satisfy $\tilde kL\equiv kL$ mod $2\pi$.
\section{Low density limit of the $v_i=0$ system}
\label{app:lowd}
Let us consider a homogeneous system $v_i=0$ with only two electrons.
These electrons form either singlet ($S=0$) or triplet ($S=1$).
For the triplet, the two electrons suffer Pauli blocking effect
and the Coulomb energy becomes exactly zero,
which means that the total energy is
exactly $E(S=1)=\epsilon(k_1)+\epsilon(k_2)$, where $\epsilon(k)=-2t\cos k$.
\par
The singlet energy eigenstates have the form,
\begin{equation}
    |\Psi\rangle = \sum_{k_1,k_2}\delta_{k_1+k_2,k} F(k_1, k_2)
        c^\dag_{k_1,\uparrow} c^\dag_{k_2,\downarrow} |0\rangle
\end{equation}
with $F(k_2, k_1) = F(k_1, k_2)$.
Plugging it into the eigenequation, $(\mathcal{H}-E)|\Psi \rangle=0$, we find,
\begin{align}
&\delta'_{k_1+k_2,k}\bigg((\epsilon(k_1)+\epsilon(k_2)-E) F(k_1,k_2)
\notag\\
&\rule{10mm}{0mm}
   +\frac{U}{N_s} \sum_{k_3, k_4}\delta'_{k_3 + k_4,k}F(k_3, k_4)\bigg)= 0
\end{align}
where $\delta'_{k_1,k_2}=1$ for $k_1\equiv k_2$ mod $2\pi$ and otherwise zero.
We further transform it into
\begin{align}
&    -\frac{1}{U}\delta_{k_1 + k_2,k}F(k_1, k_2)
\notag\\
&=\frac{1}{N_s}\frac{\delta_{k_1 + k_2, k}}{\epsilon(k_1)+\epsilon(k_2)-E}\sum_{k_3, k_4}\delta_{k_3 + k_4, k}F(k_3, k_4)\label{eq:2e_wave_func}
\end{align}
Taking the summation over $k_1$, $k_2$ in both sides and dividing them with $\sum_{k_1,k_2}\delta_{k_1 + k_2,Q}F(k_1, k_2)$, we obtain,
\begin{equation}
g(E)=-\frac{1}{U}
\label{eq:ge}
\end{equation}
with
\begin{equation}
g(E)=\dfrac{1}{N_\mathrm{s}}\sum_{k_1, k_2}\frac{\delta_{k_1 + k_2, k}}{\epsilon_{k_1} + \epsilon_{k_2}-E}=\dfrac 1{N_\mathrm{s}}\sum_{n=1}^M\dfrac{w_n}{E_n^{(0)}-E}.
\end{equation}
For a fixed total momentum, $k=k_1+k_2$, we sort the values of $\epsilon(k_1)+\epsilon(k_2)$ in ascending order, name them as $E_n^{(0)}$ ($n=1,2,\cdots,M$), and define $w_n$ as their degeneracy numbers.
Because, $\lim_{E\to E_n^{(0)}\mp 0}g(E)=\pm\infty$ the solution for Eq.(\ref{eq:2electron})
lies in the range of $E_n^{(0)}<E<E_{n+1}^{(0)}$, which form an energy continuum in $|E|\le |4t\cos(k/2)|$ for $N_\mathrm{s}\to\infty$.
\par
On the other hand, since $\lim_{E\to E_M^{(0)}+0} g(E)=-\infty$ and
$\lim_{E\to+\infty}g(E)=0$, there is a solution at $E>|4t\cos(k/2)|$,
which form the two-electron bound state.
In this energy range, we find
\begin{align}
g(E)&=\dfrac{1}{N_\mathrm{s}}\sum_{k_1, k_2}\frac{\delta_{k_1 + k_2, k}}{\epsilon(k_1)+\epsilon(k_2)-E}\notag\\
&\stackrel{N_\mathrm{s}\rightarrow+\infty}\to\dfrac 1{2\pi}\int_{-\pi}^{+\pi} dk_1\dfrac 1{-4t\cos(k/2)\cos(k_1-k/2)-E}\notag\\
&=-\dfrac 1{\sqrt{E^2-(4t\cos(k/2))^2}},
\end{align}
and obtain the bound state energy as
\begin{equation}
    E=E(k)=\sqrt{U^2+\left(4t\cos\dfrac k2\right)^2}.
\label{eq:doublon}
\end{equation}
At $U\gg t$,
$E(k)\approx U+\dfrac{8t^2}U\cos^2\dfrac k2=U+\dfrac{4t^2}U(1+\cos k)$,
so that the dispersion becomes flattened.
\par
Considering the ground state in the dilute limit, $\mu\to\epsilon(k=0)=-2t$, 
one only needs to consider the excitation that adds one electron. 
The two-electron bound state manifests in the peak at
\begin{equation}
    \omega=E(k)+2t-\mu
   =\left(U^2+\left(4t\cos\dfrac k2\right)^2\right)^{1/2}+2t-\mu. 
\label{eq:2electron}
\end{equation}

\bibliography{rscptref}

\end{document}